\documentclass[%
 reprint,
 amsmath,amssymb,
 aps,
 prl,
]{revtex4-1}

\newcommand{\nn}{\nonumber}

\newcommand{\plm}{P_l^m}

\newcommand{\hl}{h_l}

\newcommand{\rb}{\bar{r}}
\newcommand{\rhb}{\bar{\rho}}
\newcommand{\zb}{\bar{z}}

\usepackage{graphicx}
\usepackage{dcolumn}
\usepackage{bm}

\begin{document}

\title{Folding a focalized acoustical vortex on a flat holographic transducer: miniaturized selective acoustical tweezers.}

\author{Michael Baudoin}
\email{Corresponding author: michael.baudoin@univ-lille1.fr}
\homepage{http://films-lab.univ-lille1.fr/michael/michael/home.html}
\affiliation{Univ. Lille, CNRS, Centrale Lille, ISEN, Univ. Valenciennes, UMR 8520 - IEMN, International laboratory LIA/LICS, F-59000 Lille, France}%
\author{Jean-Claude Gerbedoen}%
\affiliation{Univ. Lille, CNRS, Centrale Lille, ISEN, Univ. Valenciennes, UMR 8520 - IEMN, International laboratory LIA/LICS, F-59000 Lille, France}%
\author{Antoine Riaud}
\affiliation{Universit\'{e} Paris Sorbonne Cit\'{e}, INSERM UMR-S1147, 45 rue des Saints-P\`{e}res, 75270, Paris, France}%
\altaffiliation{Current affiliation: ASMC and System State Key Laboratory, School of Microelectronics, Fudan University, Shanghai 200433, China}
\author{Olivier Bou Matar}%
\affiliation{Univ. Lille, CNRS, Centrale Lille, ISEN, Univ. Valenciennes, UMR 8520 - IEMN, International laboratory LIA/LICS, F-59000 Lille, France}%
\author{Nikolay Smagin}%
\affiliation{Univ. Lille, CNRS, Centrale Lille, ISEN, Univ. Valenciennes, UMR 8520 - IEMN, International laboratory LIA/LICS, F-59000 Lille, France}%
\author{Jean-Louis Thomas}
\affiliation{Sorbonne Universit\'{e}s, UPMC Univ Paris 06, CNRS UMR 7588, Institut des NanoSciences de Paris, 4 place Jussieu, 75005 Paris, France}%

\date{\today}


\begin{abstract}
Acoustical tweezers based on focalized acoustical vortices hold the promise of precise contactless 3D manipulation of millimeter down to sub-micrometer particles, microorganisms and cells with unprecedented combined selectivity and trapping force. Yet, the widespread dissemination of this technology has been hindered by severe limitations of current systems in terms of performance and/or miniaturization and integrability. In this paper, we unleash the potential of focalized acoustical vortices by developing the first flat, compact, single-electrodes focalized acoustical tweezers. These tweezers rely on holographic Archimedes-Fermat spiraling transducers obtained by folding  a spherical acoustical vortex on a flat piezoelectric substrate. We demonstrate the ability of these tweezers to grab and displace micrometric objects in a standard microfluidic environment with unique selectivity. The simplicity of this system and its scalability to higher frequencies opens tremendous perspectives in microbiology, microrobotics and microscopy.

\end{abstract}

\pacs{Valid PACS appear here}
\maketitle


\section{\label{sec:method}Introduction}

The precise contactless manipulation of physical and biological objects at micrometric down to nanometric scales promises tremendous development in fields as diverse as microrobotics, tissue engineering or micro/nano-medicine. In this regard, acoustical tweezers is a prominent technology since it is non-invasive, biocompatible \cite{hultstrom2007,wiklund2012,po_burguillos_2013}, label-free and enables trapping forces several orders of magnitudes larger than their optical counterparts at same actuation power \cite{jasa_baresch_2013,baresch2016}. First reported observations of particles levitation in acoustic wavefields date back to the work of Boyle and Lehmann \cite{trsc_boyle_1925}. Nevertheless, only recent simultaneous developments of advanced wave synthesis systems, microfluidic setups and theory of acoustic radiation pressure enabled to harness the potential of acoustophoresis \cite{shi2009,tran2012,ding2012,pre_Barnkob_2012,pnas_poulikakos_2013,ding2014,courtney2014,po_ochiai_2014,marzo2015,collins2015b,natcom_augustsson_2016,baresch2016,loc_neild_2017,prap_riaud_2017}. Until recently, the vast majority of acoustical tweezers relied on a single or a set of orthogonal standing waves that create a network of nodes and antinodes where particles are trapped \cite{shi2009,tran2012,ding2012,pre_Barnkob_2012,pnas_poulikakos_2013,ding2014,po_ochiai_2014,collins2015b,natcom_augustsson_2016,loc_neild_2017}. These systems are highly efficient for the collective manipulation of particles and cells but the multiplicity of traps and the agglomeration of several particle at the same node or antinode  preclude any selectivity. While limited localization of the acoustic energy can be achieved by the original sub-time-of-flight technics \cite{sa_collins_2016}, only strong focalization of wavefields enables selectively at the single particle level. Focalized acoustic waves are natural candidates to achieve this localization \cite{lee2009,li2013,hwang2016}, but many particles of practical interest migrate to the nodes of standing wavefields like rigid particles and cells \cite{gorkov1962,settnes2012} and are thus expelled from the wave focus. This radial expelling and the condition to get a restoring axial force are the main difficulties that thwarted research on selective acoustical tweezers.  The combination of strong localization and the existence of a minimum of the pressure wavefield at the focus point surrounded by a bright ring is fulfilled by so-called cylindrical or spherical acoustical vortices, some helical waves spinning around a phase singularity \cite{prsa_nye_1974,hefner1999,thomas2003,jap_barech_2013}. These waves are the separate variable solutions of Helmholtz equation in cylindrical and spherical coordinates respectively. The former are invariant in the z-direction and thus enable only 2D particle trapping, while focalized spherical vortices enable 3D particle trapping with a single beam, as first demonstrated theoretically by Baresch et al. \cite{jasa_baresch_2013,jap_barech_2013}, following recent developments in the theory of acoustic radiation force \cite{marston2006,jasa_marston_2009}. A wealth of systems \cite{thomas2003,courtney2014,marzo2015,riaud2015a,riaud2015b,Riaud2016,baresch2016,prl_jiang_2016,pre_jimenez_2016,mupb_terzi_2017,prap_riaud_2017} have been proposed since the seminal work by Marston \cite{hefner1999} for the synthesis of acoustical vortices. Nevertheless the ability to obtain a 3D trap and to pick up one particle independently  of its neighbors was only demonstrated recently by Baresch et al \cite{baresch2016} .This operation requires a strong focalization of the acoustical vortex \cite{jap_barech_2013}. In addition, all acoustical vortices synthesis systems developed up to date rely on either arrays of transducers \cite{thomas2003,courtney2014,marzo2015,riaud2015a,riaud2015b,baresch2016}  or passive systems \cite{prl_jiang_2016,pre_jimenez_2016,mupb_terzi_2017} that are cumbersome, hardly miniaturizable and not compatible with microscopes or microfluidics chips. Recently, Riaud et al. \cite{prap_riaud_2017} showed that it is possible to generate cylindrical acoustical vortices with swirling surface acoustic waves synthesized with spiraling InterDigitated Transducers (IDTs) printed at the surface of a piezoelectric substrate. This system is flat, transparent, compatible with disposable substrates and its fabrication with standard lithography is straightforward to miniaturize and cheap as proved by its widespread use in modern electronic devices \cite{mssp_emanetoglu_1999}. This wave synthesis system nevertheless suffers from major limitations:  (i) lateral focalization of cylindrical acoustical vortices is weaker than 3D focalization of their spherical counterparts and leads to the existence of spurious secondary rings of weaker amplitude (see Fig. 2.e) that can also trap particles, hence severely limiting the tweezers selectivity, (ii) the anisotropy of piezoelectric substrates leads to an anisotropy of surface waves and hence of the trapping force, (iii) 3D trapping is not possible with cylindrical acoustical vortices. In summary, the swirling SAW technology has come with its own set of challenges and limitations, and a paradigm shift is needed to reach true selectivity without compromising on miniaturization.

In this paper,  we harness the potential of selective acoustical tweezers by folding an isotropic focalized acoustical vortex on a flat surface and synthesize it with single spiraling interdigitated electrodes reminiscent of a Archimedes-Fermat spiral. The radial contraction of the spiral enables wave focusing without the requirement of a curved transducer or a lens, a major advantage compared to existing systems. In addition the limitations of the cylindrical vortex based tweezers are overcome. We demonstrate the high selectivity of this tweezer (i) by measuring the acoustic field with a laser interferometer and quantifying the fast radial decrease of secondary rings and (ii) by selectively trapping and moving one particle independently of its neighbors in a standard microfluidic environment. 

\onecolumngrid
\begin{center}
\begin{figure}[htbp]
\includegraphics[width=\textwidth]{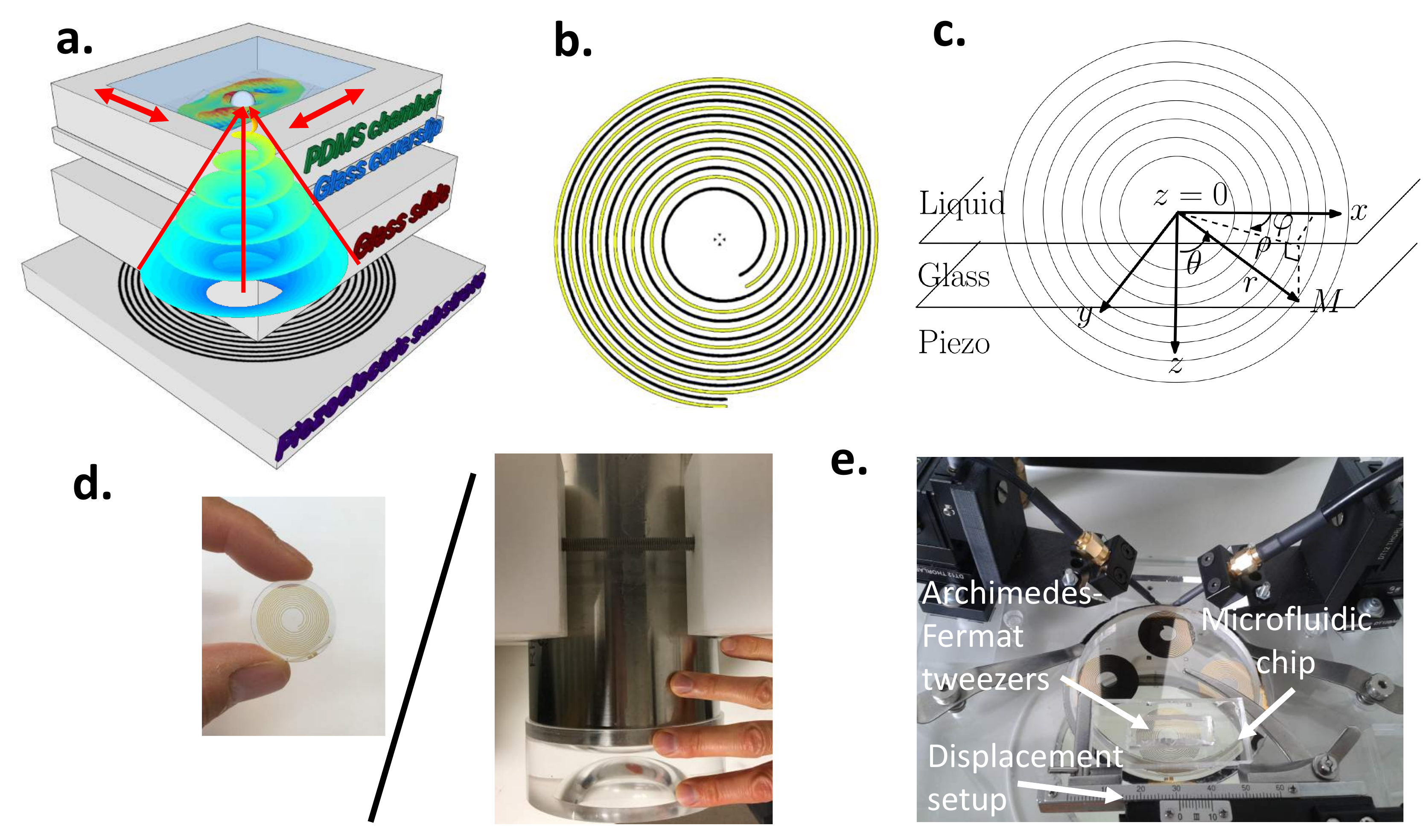}
\caption{a. Scheme illustrating the composition of the Archimedes-Fermat acoustical tweezers: A focalized acoustical vortex is synthesized by spiraling metallic electrodes sputtered at the surface of a piezoelectric substrate. The vortex propagates and focalizes inside a glass slide (sealed with the piezoelectric substrate) and a mobile glass coverslip before reaching the liquid contained in a PDMS chamber, wherein the particle is trapped. The mobility of the microfluidic chip (glass coverslip + sealed PDMS chamber) is enabled by a liquid couplant and a manual precision displacement setup represented on Fig.e.. b. Spiraling pattern of the electrodes obtained from approximate equation (\ref{eq:spiral}) with $\bar{z} = 55.4$ and m = 1. c. Scheme introducing the spherical $(r, \, \theta, \, \varphi)$ and cylindrical coordinates $(\rho, \,\varphi, \, z)$ used for the demonstration of equation (\ref{eq:spiral}). d. Comparison of the compactness of the Archimedes-Fermat acoustical tweezers presented in this paper (left) to the transducer array of ref. \cite{baresch2016} (right). This figure also shows the transparency of the Archimeded-Fermat acoustical tweezers (particles are trapped on the central axis of the transducer). e. Picture showing the integration of the Archimedes-Fermat acoustical tweezers  into a Leica Z16 macroscope. Four tweezers have been patterned on a 3 inch LiNb03 wafer.}   
\label{figure1}
\end{figure}
\end{center}
\twocolumngrid

\section{Materials and methods}

\subsection{\label{sec:method_theory}Theory: Folding a spherical acoustical vortex on a flat surface.}

Spherical acoustical Bessel beams (spherical vortices) constitute excellent candidates to create localized acoustic trap. Indeed, these acoustic fields both focalize the acoustic energy in 3D and create a shadow zone at the center of the vortex surrounded by a bright shell, wherein particles can be trapped \cite{jap_barech_2013}. In the same way that a plane standing wave can be seen as the combination of two counter-propagating traveling waves, a spherical Bessel beam results from the interference  between a converging and a diverging spherical Hankel beam (of the first and second kind respectively). Hence a Bessel beam can be produced by a single Hankel converging beam, which will interfere with its diverging counterpart generated at the focus, i.e. at the vortex central singularity. In this paper, we demonstrate that  converging Hankel beams can be synthesized by materializing the intersections between some isophases surfaces and a plane, with metallic electrodes sputtered at the surface of a piezoelectric substrate  (see Fig. \ref{figure1} a.). Indeed, the electric excitation of these electrodes provokes localized vibrations of the piezoelectric substrate, which in turn produce a bulk acoustic vortex inside a glass slide. This original holographic method combines the underlying physical principles of Fresnel lenses in optics (wherein an isophase is folded on a flat surface), the specificity of Bessel beams topology and the principles of wave synthesis with IDTs in the field of microelectronics. 

The general expression of a Hankel spherical vortex in the complex plane is given by $$\Psi^*(r,\theta,\varphi) = A \hl \plm (\cos(\theta)) e^{i (-m \varphi + \omega t)},$$ with $r$, $\theta$, $\varphi$ the spherical coordinates, $k = \omega / c_a$ the wave number, $c_a$ the sound speed, $A>0$ the wave amplitude, $\plm$ the Legendre polynomial of order ($l$, $m$), $\omega$ the frequency, $t$ the time and $\hl$ the spherical Hankel function of the first kind. To compute the intersection of the Hankel beam with a plane, we introduce the cylindrical coordinates $\rho$, $\varphi$, $z$, with: $r = \sqrt{\rho^2 + z^2}$ and $z$ the axis normal to the plane (see Fig. \ref{figure1}.c). For the sake of simplicity we also introduce the dimensionless parameters $\rb = kr$, $\rhb = k \rho$, $\zb = kz$. In the far field ($\bar{z} \gg 1$ and thus $\bar{r} \gg 1$), Hankel functions of the first kind can be approximated by $\hl(\bar{r}) = (-i)^{l+1} e^{i\bar{r}} / \bar{r}$ and thus isophases of $\Psi^*$ are simply given by the equation:
\begin{eqnarray}
\phi & = & \arg(\Psi^*) = \arg(\hl(\rb)) + \arg(e^{i (-m \varphi + \omega t)}) \label{eq:spirale} \\
& \approx & \rb - (l+1) \frac{\pi}{2} - m \varphi + \omega t = C_1 \nn
\end{eqnarray}
with $C_1$ a constant.
Since $t$ and the constant (left hand side) are arbitrary chosen, this equation reduces to the simple expression: $\phi =  \rb - m \varphi = C_2$, with $\bar{r} = \sqrt{\bar{\rho}^2 + \bar{z}^2}$ and $C_2$ a constant. 

Because of the piezoelectric effect, the mechanical vibrations of the bulk acoustic waves are coupled to the electrical potential. Following  the principle of wave synthesis with IDTs, we model the electrodes as perfect wires (isopotential lines). If we consider two spirals in opposition of phase ($\Delta \phi = \pi$), we obtain the following polar equations of the metallic tracks:
\begin{equation}
\rhb = \sqrt{(m \varphi + C_2)^2 - \zb^2 }  \mbox{ and }  \rhb =  \sqrt{(m \varphi + C_2 + \pi)^2 - \zb^2 }
\label{eq:spiral}
\end{equation}

These simple equations correspond to spirals, as represented on Fig. \ref{figure1}.b (with $\bar{z} = 55.4$, m = 1). It is an intermediate situation between the Fermat spiral ($\rhb\propto \varphi^2$) and the Archimedes spiral ($\rhb\propto \varphi$). The radial distance between consecutive elecrodes decreases progressively (as shown by the second derivative of these equations) and the spirals converge asymptotically toward Archimedes arithmetic spirals of equation $\rhb = m \varphi + K$ (with $K=0$ or $\pi$). This geometric regression of the radial distance between consecutive tracks ensures the beam focusing. A pure Archimedes spiral would generate a cylindrical vortex invariant in the $z$ direction, while a pure Fermat spiral would require an infinitely thick substrate. We can note that equation (\ref{eq:spiral}) is only a far-field approximation of the exact solution of the problem. The exact shape of the spirals can be obtained by solving directly the nonlinear equation (\ref{eq:spirale}). We solved numerically this equation with an interior-point method and observed that the approximate and exact spirals only differ in the vicinity of the spiral center (see Image I1 in SM). Since the center of the spiral is not materialized in the present tweezers (to leave space for visualization), this difference does not affect the synthesized wavefield.

\subsection{\label{sec:method_exp}Experimental design of the tweezer.}

Experimentally,  a system was designed to synthesize focalized vortices of topological order $(l,m) = (1,1)$ at frequency $f = \omega/2\pi =  4.4$ MHz. This system relies on spiraling metallic electrodes sputtered at the surface of a $0.5$mm thick Y-$36$ Niobate Lithium (LiNbO3) piezoelectric substrate following a standard lift-off procedure: (i) a negative mask with the electrodes represented on Fig. \ref{figure1}.b is obtained with high resolution printing, (ii) a sacrificial layer (photoresist AZnLOF2020) of thickness $3 \; \mu$m is deposited on the substrate and patterned using conventional photholithography technique,  (iv) a titanium (Ti) layer of 20 nm and a gold (Au) layer of 200 nm are successively sputtered on the LiNbO3 substrate and (v) the sacrificial layer is washed out with N-Methyl-2-pyrrolidone (NMP) based solvent stripper. The vibration of these spiraling electrodes driven by a TEKTRONIX  AFG3051C waveform generator and a AR150A250 amplifier instills a converging Hankel beam inside a borofloat glass slide of $6.5$ mm thickness glued at the surface of the piezoelectric substrate with EPO-TEK 301-2 (see Fig. \ref{figure1}a). The properties of the glass slide is chosen to match the acoustic properties of a mobile borofloat glass coverslip of $150 \, \mu$m thickness supporting a PDMS channel of $300$ $\mu$m depth glued with $O_2$ plasma. The microfluidic chip (glass coverslip + PDMS chamber) containing the particles is coupled with the glass slide with a drop of oil and moved with a manual precision displacement setup (Fig. \ref{figure1}e.). The electrodes are designed (i) to generate transverse waves of speed $c_T$ into the glass slide, (ii) to obtain a focalization at the surface of the coverslip and thus enable optimal wave localization in the microfluidic chamber and (iii) to obtain a transducer aperture of $64^o$. Transverse waves in the glass were chosen owing to their lower sound speed ($c_T = 3280 m s^{-1}$ ), which ensures better transmission of the acoustic energy from the glass to the liquid. The resulting acoustic field at the surface of the glass coverslip (focal plane, z=0) was measured with a UHF-120 Polytec laser vibrometer

\onecolumngrid
\begin{center}
\begin{figure}[htbp]
\includegraphics[width=\textwidth]{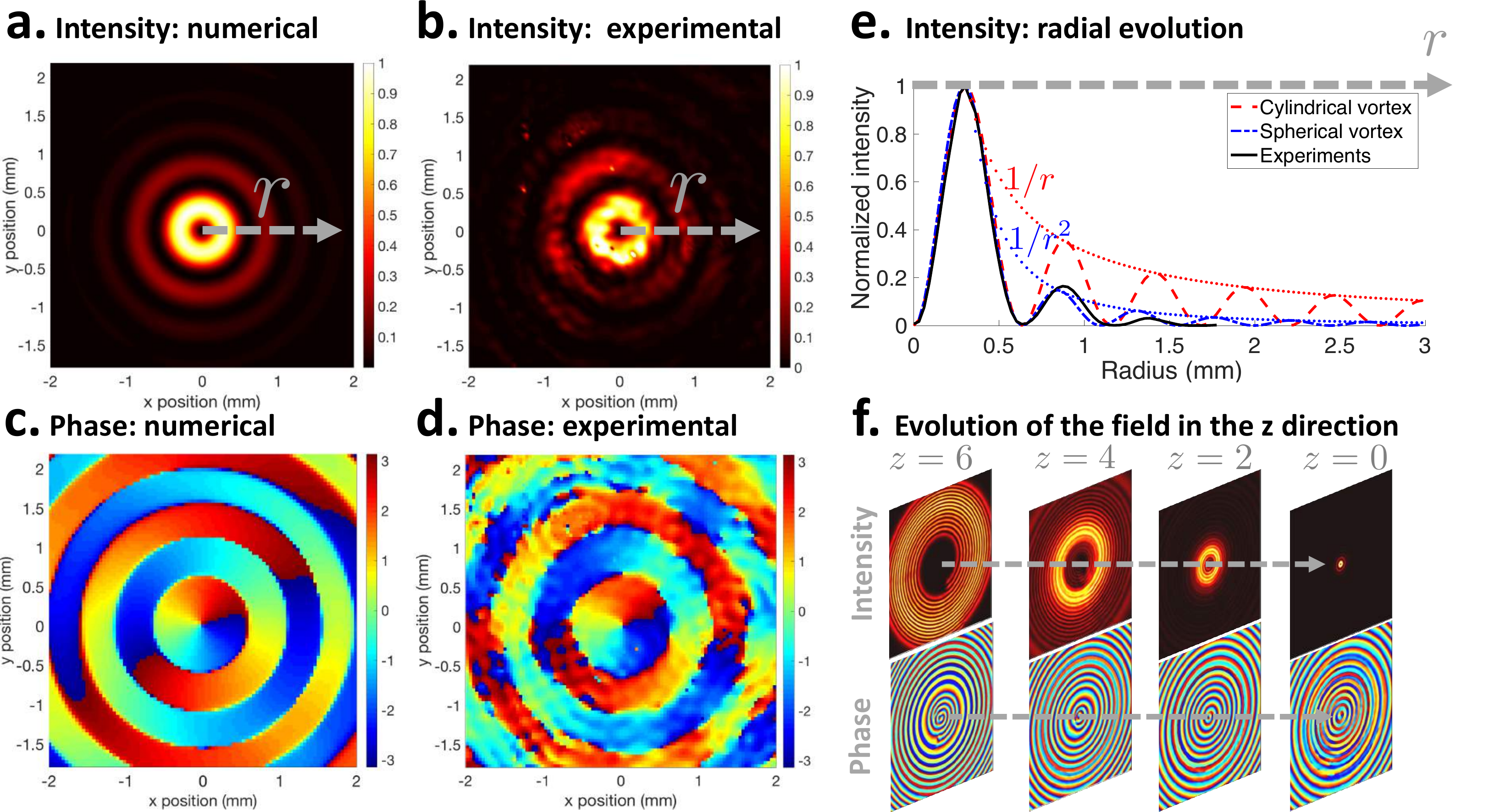}
\caption{a. Numerical predictions  with angular spectrum and b. experimental measurements with a UHF-120 Polytec laser interferometer of the normalized intensity of the vibration at the surface of the glass coverslip (focal plane, $(z=0)$). The maximum amplitude measured experimentally (on the first ring) is $10$nm.  c.Numerical predictions  with angular spectrum and d. experimental measurements with the laser interferometer of the phase of the acoustic wave at the surface of the glass coverslip. e. Radial evolution of the normalized intensity of the acoustic wave from the center of the vortex to the side, as a function of the lateral radius $r$ in mm. Black solid line: average over all angles $\varphi$ of the intensity measured experimentally.  Red dashed line: evolution expected for a cylindrical vortex (cylindrical Bessel function). Blue dashed-dotted line: evolution expected for a spherical vortex (spherical Bessel). Red dotted line: asymptotic evolution in $1/r$. Blue dotted line: asymptotic evolution in  $1/r^2$. f. Evolution of the field intensity (up) and phase (down) in the z direction. The direction of the arrow indicates the wave propagation direction. Left to right: distances $z = 6, 4, 2, 0$ mm respectively ($z=0$ corresponds to the focal plane). The upper part of the figure shows the focalization of the acoustic energy and the formation of a localized trap. The lower part shows the transition from a Hankel to a Bessel spherical beam.}   
\label{figure2}
\end{figure}
\end{center}
\twocolumngrid

\subsection{\label{sec:method_num} Numerical prediction of the resulting acoustic field.}

Predictions of the acoustic field expected at the surface of the glass coverslip (in contact with the liquid) were obtained by propagating the vibrations produced by the spiraling transducers in the glass slide and coverslip with an angular spectrum code \cite{piee_Booker_1950}. This latter method relies on (i) the 2D spatial Fourier transform of a wave field in a source plane (electrode plane), (ii) the propagation of each corresponding plane wave in the Fourier space to a target plane and (iii) the inverse Fourier transform of the result. 

\onecolumngrid
\begin{center}
\begin{figure*}[htbp]
\includegraphics[width=\textwidth]{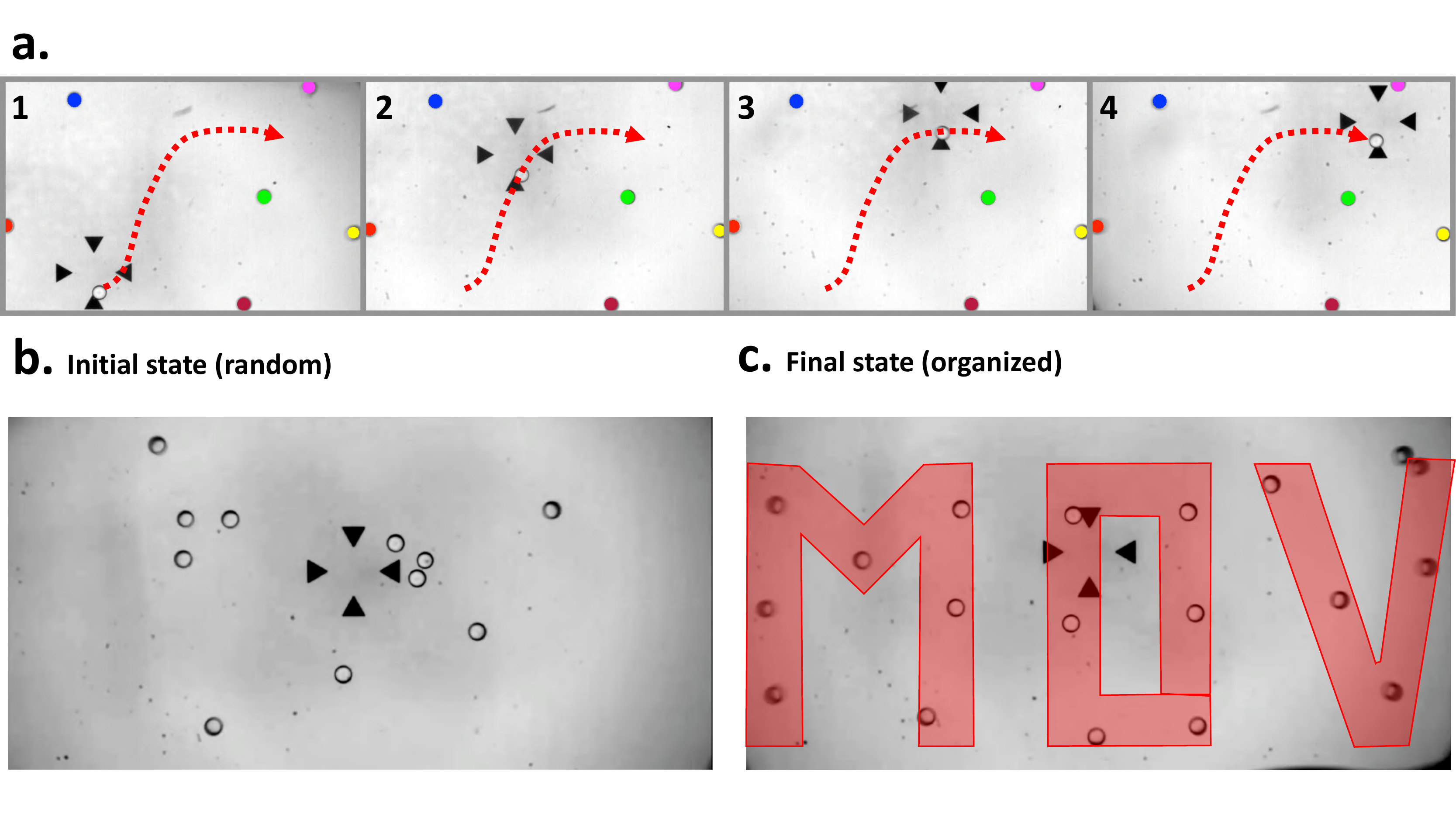}
\caption{a. Selective manipulation of a 75 $\pm  2 \mu$m radius polystyrene particle with the $4.4$ MHz selective acoustical tweezers based on Archimedes-Fermat spirals. This figure shows that only the particle trapped at the center of the vortex (located just above the lowest arrow as shown on movie M3 in SM) is moved while the other particles remain still. The particles at rest have been colored to improve the readability of the figure. See also supplementary movie M2 in SM. b.c. Patterning of 18 polystyrene particles (radius 75 $\pm  2 \mu$m) into prescribed position to form the letters MOV (Moving Object with Vortices). b. Initial state: particles randomly dispersed. c. Final states: organized particles.}   
\label{figure3}
\end{figure*}
\end{center}
\twocolumngrid

\section{\label{sec:exper}Results}

We first compared the acoustic field measured experimentally to numerical predictions obtained from angular spectrum  (see Fig. \ref{figure2}a-d and movie M1 in SM). Excellent agreement between numerical predictions and experiments is obtained for both the intensity (Fig. \ref{figure2}.a and b) and the phase (Fig. \ref{figure2}.c and d) of the wavefield. In addition, the radial evolution of the rings intensity (in the ($x$, $y)$ plane) measured experimentally and averaged over all angles $\varphi$ (black continuous line) is compared to (i) the radial evolution of a cylindrical vortex (red dashed line) and (ii) the radial evolution of a spherical vortex (blue dashed-dotted line  on Fig. \ref{figure2}e). This figure shows (i) that the radial evolution of the vortex intensity measured experimentally closely follows the one of a spherical vortex and (ii) that the intensity of secondary rings decreases much faster for spherical vortices ($\propto 1/r^2$) than for cylindrical vortices  ($\propto 1/r$). Since radiation pressure is proportional to the intensity of the beam, it means that the selectivity is greatly enhanced by the axial focusing of the beam, compared to cylindrical vortices. Thus 3D focalization of the energy is a major advantage for selective manipulation of particles. The slight differences between the experimentally measured field and the spherical Bessel function can be explained by the fact that we do not literally synthesize a spherical Bessel vortex. Indeed the phase of our signal matches the one of a spherical vortex but not the radial evolution of the amplitude.

To demonstrate the selectivity of this acoustical tweezer, i.e. its ability to pick up one particle and move it independently of its neighbors, we dispersed some monodisperse polystyrene particles of $75 \pm 2 \; \mu m$ radius inside a microfluidic chamber of 300 microns height  and then picked one particle and moved it around between the other particles. The power of our beam was chosen to produce a sufficient force to move the target particle while keeping the second ring magnitude sufficiently weak not to be able to trap particles. Fig. \ref{figure3}.a and movie M2 in SM clearly show that as long as the distance between the transported particle and the other particles remains larger than the radius of the first ring, only the trapped particles moves. Of course, when the first ring touches a particle located outside it, then the particle is pushed away since the rings are repulsive. It is important to note that, in this video, the center of the vortex is located at the tip of the bottom arrow where the particle is trapped (as demonstrated by the scan of the acoustic field, Movie M3 in SM). We additionally demonstrated the tweezers ability to precisely position a set of 18 polystyrene particles of $75 \pm 2 \; \mu m$ radius in a prescribed pattern (MOV: Moving Objects with Vortices), starting from randomly distributed particles (see Fig. \ref{figure3}b-c).

\section{\label{sec:exper}Discussion}

Acoustic tweezers have long faced a trade-off between selectivity and miniaturization/integrability. These restrictions have so far prevented the development of many applications in microfluidics and microbiology. In this paper, these limitations are overcome by unifying the physical principles of (i) acoustic trapping with spherical vortices, (ii) holographic wave synthesis with interdigitated transducers and (iii) Fresnel lenses, inside a single compact transparent miniaturized device. With this microsystem, we demonstrate the contactless manipulation of particles in a standard microscopy environment with state of the art selectivity. Owing to the simplicity of the technology and its scalability to higher frequencies, this work paves the way towards individual manipulation and in-situ assembly of physical and biological micro-objects.

\section{Supplementary materials} Supplementary material for this article is available at ... \\
Image S1: Comparison of the shape of the electrodes obtained by approximated equation (\ref{eq:spiral}) and exact equation (\ref{eq:spirale}). \\
Movie M1: Movie showing an animation of the vortex measured experimentally with the laser interferometer.\\
Movie M2: Movie showing the selective manipulation of a $75 \pm 2 \mu m$ radius polystyrene particle with the $4.4$ MHz selective acoustical tweezers based on Archimedes-Fermat spirals.\\
Movie M3: Movie showing the localization of the vortex core compared to the localization of the arrows. \\

\section{Additional information}

\noindent \textbf{Acknowledgements:} We acknowledge Alexis Vlandas for his critical reading of the manuscript and valuable advices that helped us improve its overall quality.  \noindent \textbf{Funding:} This work was supported by SATT du Nord. \noindent \textbf{Authors contributions:} M.B., A.R. and J.-L.T. designed research; M.B. and J.-C.G. built the setup and performed the experiments; M.B. and A.R. performed the numerical simulations; O. B.M., J.-C..G. and N.S. performed the field measurements with the interferometer; M.B., O. B.M., A.R. and J.-L.T. analyzed and interpreted the results; M.B., O. B.M., J.-C. G., A.R, J.-L.T. wrote the paper. All authors approved the final version of the manuscript. \textbf{Competing interests:} The authors declare no competing interests.  \noindent \textbf{Data and materials availability: } All data needed to evaluate the conclusions in the paper are present in the paper and/or the Supplementary Materials. Additional data related to this paper may be requested from the authors.

\bibliographystyle{ScienceAdvances}


\begin{thebibliography}{10}

\bibitem{hultstrom2007}
J.~Hultstr{\"o}m, O.~Manneberg, K.~Dopf, H.~M. Hertz, H.~Brismar, M.~Wiklund,
  Proliferation and viability of adherent cells manipulated by standing-wave
  ultrasound in a microfluidic chip.
\newblock {\it Ultras. Med. Biol.\/} {\bf 33}, 145--151 (2007).

\bibitem{wiklund2012}
M.~Wiklund, Acoustofluidics 12: Biocompatibility and cell viability in
  microfluidic acoustic resonators.
\newblock {\it Lab Chip\/} {\bf 12}, 2018--2028 (2012).

\bibitem{po_burguillos_2013}
M.~Burguillos, C.~Magnusson, M.~Nordin, A.~Lenshof, P.~Austsson, M.~J. Hansson,
  E.~Elmér, H.~Lilja, P.~Brundin, Y.~Laurell, T.~Deierborg, Microchannel
  acoustophoresis does not impact survival or function of microglia, leukocytes
  or tumor cells.
\newblock {\it PloS ONE\/} {\bf 8}, e64233 (2013).

\bibitem{jasa_baresch_2013}
D.~Baresch, J.~Thomas, R.~Marchiano, Three-dimensional acoustic radiation force
  on an arbitrarily located elastic sphere.
\newblock {\it J. Acoust. Soc. Am.\/} {\bf 133}, 25 (2013).

\bibitem{baresch2016}
D.~Baresch, J.-L. Thomas, R.~Marchiano, Observation of a single-beam gradient
  force acoustical trap for elastic particles: acoustical tweezers.
\newblock {\it Phys. Rev. Lett.\/} {\bf 116}, 024301 (2016).

\bibitem{trsc_boyle_1925}
R.~Boyle, J.~Lehmann, A new photographic method to demonstrate the interference
  of longitundinal wave trains. the velocity of high frequency sound in a
  liquid.
\newblock {\it Trans. Roy. Soc. Canada\/} {\bf 19}, 159-165 (1925).

\bibitem{shi2009}
J.~Shi, D.~Ahmed, X.~Mao, S.-C.~S. Lin, A.~Lawit, T.~J. Huang, Acoustic
  tweezers: patterning cells and microparticles using standing surface acoustic
  waves (ssaw).
\newblock {\it Lab Chip\/} {\bf 9}, 2890--2895 (2009).

\bibitem{tran2012}
S.~Tran, P.~Marmottant, P.~Thibault, Fast acoustic tweezers for the
  two-dimensional manipulation of individual particles in microfluidic
  channels.
\newblock {\it Appl. Phys. Lett.\/} {\bf 101}, 114103 (2012).

\bibitem{ding2012}
X.~Ding, S.-C.~S. Lin, B.~Kiraly, H.~Yue, S.~Li, I.-K. Chiang, J.~Shi, S.~J.
  Benkovic, T.~J. Huang, On-chip manipulation of single microparticles, cells,
  and organisms using surface acoustic waves.
\newblock {\it P. Nat. Acad. Sci. USA\/} {\bf 109}, 11105--11109 (2012).

\bibitem{pre_Barnkob_2012}
R.~Barnkob, P.~Augustsson, T.~Laurell, H.~Bruus, Acoustic radiation- and
  streaming- induced microparticle velocities determined by microparticle image
  velocimetry in an ultrasound symmetry plane.
\newblock {\it Phys. Rev. E\/} {\bf 86}, 056307 (2012).

\bibitem{pnas_poulikakos_2013}
D.~Foresti, M.~Nabavi, M.~Klingauf, A.~Ferrari, D.~Poulikakos, Acoustophoretic
  contactless transport and handling of matter in air.
\newblock {\it Proc. Nat. Ac. Sci. USA\/}  (2013).

\bibitem{ding2014}
X.~Ding, Z.~Peng, S.-C.~S. Lin, M.~Geri, S.~Li, P.~Li, Y.~Chen, M.~Dao,
  S.~Suresh, T.~J. Huang, Cell separation using tilted-angle standing surface
  acoustic waves.
\newblock {\it P. Nat. Acad. Sci. USA\/} {\bf 111}, 12992--12997 (2014).

\bibitem{courtney2014}
C.~R. Courtney, C.~E. Demore, H.~Wu, A.~Grinenko, P.~D. Wilcox, S.~Cochran,
  B.~W. Drinkwater, Independent trapping and manipulation of microparticles
  using dexterous acoustic tweezers.
\newblock {\it Appl. Phys. Lett.\/} {\bf 104}, 154103 (2014).

\bibitem{po_ochiai_2014}
Y.~Ochiai, T.~Hoshi, J.~Rekimoto, Three-dimensional mid-air acoustic
  manipulation by ultrasonic phased arrays.
\newblock {\it Plos ONE\/} {\bf 9}, e97590 (2014).

\bibitem{marzo2015}
A.~Marzo, S.~A. Seah, B.~W. Drinkwater, D.~R. Sahoo, B.~Long, S.~Subramanian,
  Holographic acoustic elements for manipulation of levitated objects.
\newblock {\it Nat. Commun.\/} {\bf 6} (2015).

\bibitem{collins2015b}
D.~J. Collins, B.~Morahan, J.~Garcia-Bustos, C.~Doerig, M.~Plebanski, A.~Neild,
  Two-dimensional single-cell patterning with one cell per well driven by
  surface acoustic waves.
\newblock {\it Nat. Commun.\/} {\bf 6} (2015).

\bibitem{natcom_augustsson_2016}
P.~Augustsson, J.~Karlsen, H.-W. Su, H.~Bruus, J.~Vlodman, Iso-acoustic
  focusing of cells for size-sensitive acousto-mechanical phenotyping.
\newblock {\it Nat. Commun.\/} {\bf 7}, 11556 (2016).

\bibitem{loc_neild_2017}
J.~Ng, C.~Devendran, A.~Neild, Acoustic tweezing of particles using decaying
  opposing travelling surface acoustic waves (dotsaw).
\newblock {\it Lab Chip\/} {\bf 17} (2017).

\bibitem{prap_riaud_2017}
A.~Riaud, M.~Baudoin, O.~Bou~Matar, L.~Becerra, J.~Thomas, Selective
  manipulation of microscopic particles with precursos swirling rayleigh waves.
\newblock {\it Phys. Rev. Appl.\/} {\bf 7}, 024007 (2017).

\bibitem{sa_collins_2016}
D.~Collins, C.~Devendran, Z.~Ma, J.~Ng, A.~Neild, Y.~Ai, Acoustic tweezers via
  sub-time-of-flight regime surface acoustic waves.
\newblock {\it Sci. Adv.\/} {\bf 2}, e1600089 (2016).

\bibitem{lee2009}
J.~Lee, S.-Y. Teh, A.~Lee, H.~H. Kim, C.~Lee, K.~K. Shung, Single beam acoustic
  trapping.
\newblock {\it Appl. Phys. Lett.\/} {\bf 95}, 073701 (2009).

\bibitem{li2013}
Y.~Li, C.~Lee, K.~H. Lam, K.~K. Shung, A simple method for evaluating the
  trapping performance of acoustic tweezers.
\newblock {\it Appl. Phys. Lett.\/} {\bf 102}, 084102 (2013).

\bibitem{hwang2016}
J.~Y. Hwang, J.~Kim, J.~M. Park, C.~Lee, H.~Jung, J.~Lee, K.~K. Shung, Cell
  deformation by single-beam acoustic trapping: A promising tool for
  measurements of cell mechanics.
\newblock {\it Sci. Rep.-UK\/} {\bf 6}, 27238 (2016).

\bibitem{gorkov1962}
L.~P. Gor'Kov, On the forces acting on a small particle in an acoustical field
  in an ideal fluid.
\newblock {\it Sovi. Phys. Dokl.\/} {\bf 6}, 773 (1962).

\bibitem{settnes2012}
M.~Settnes, H.~Bruus, Forces acting on a small particle in an acoustical field
  in a viscous fluid.
\newblock {\it Phys. Rev. E\/} {\bf 85}, 016327 (2012).

\bibitem{prsa_nye_1974}
J.~Nye, M.~Berry, Dislocations in wave trains.
\newblock {\it Proc. Roy. Soc. A\/} pp. 165--190 (1974).

\bibitem{hefner1999}
B.~T. Hefner, P.~L. Marston, An acoustical helicoidal wave transducer with
  applications for the alignment of ultrasonic and underwater systems.
\newblock {\it J. Acoust. Soc. Am.\/} {\bf 106}, 3313--3316 (1999).

\bibitem{thomas2003}
J.-L. Thomas, R.~Marchiano, Pseudo angular momentum and topological charge
  conservation for nonlinear acoustical vortices.
\newblock {\it Phys. Rev. Lett.\/} {\bf 91}, 244302 (2003).

\bibitem{jap_barech_2013}
D.~Baresch, J.-L. Thomas, R.~Marchiano, Spherical vortex beams of high radial
  degree for enhanced single-beam tweezers.
\newblock {\it J. Appl. Phys.\/} {\bf 113}, 184901 (2013).

\bibitem{marston2006}
P.~L. Marston, Axial radiation force of a bessel beam on a sphere and direction
  reversal of the force.
\newblock {\it J. Acoust. Soc. Am.\/} {\bf 120}, 3518--3524 (2006).

\bibitem{jasa_marston_2009}
P.~Marston, Radiation force of a helicoidal bessel beam on a sphere.
\newblock {\it J. Acoust. Soc . Am.\/} {\bf 125}, 3539 (2009).

\bibitem{riaud2015a}
A.~Riaud, J.-L. Thomas, M.~Baudoin, O.~Bou~Matar, Taming the degeneration of
  bessel beams at an anisotropic-isotropic interface: Toward three-dimensional
  control of confined vortical waves.
\newblock {\it Phys. Rev. E\/} {\bf 92}, 063201 (2015).

\bibitem{riaud2015b}
A.~Riaud, J.-L. Thomas, E.~Charron, A.~Bussonni{\`e}re, O.~Bou~Matar,
  M.~Baudoin, Anisotropic swirling surface acoustic waves from inverse
  filtering for on-chip generation of acoustic vortices.
\newblock {\it Phys. Rev. Appl.\/} {\bf 4}, 034004 (2015).

\bibitem{Riaud2016}
A.~Riaud, M.~Baudoin, J.~Thomas, O.~Bou~Matar, Saw synthesis with idts array
  and the inverse filter: toward a versatile saw toolbox for microfluidic and
  biological applications.
\newblock {\it IEEE T. Ultrason. Ferr.\/}  (2016).

\bibitem{prl_jiang_2016}
X.~Jiang, Y.~Li, B.~Liang, J.-C. Cheng, L.~Zhang, Convert acoustic resonances
  to orbital angular momentum.
\newblock {\it Phys. Rev. Lett.\/} {\bf 117} (2016).

\bibitem{pre_jimenez_2016}
N.~Jimenez, R.~Pico, V.~Sanchez-Morcillo, V.~Romero-Garcia, L.~Garcia-Raffi,
  K.~Staliunas, Formation of high order acoustic bessel beams by spiral
  diffraction gratings.
\newblock {\it Phys. Rev. E\/} {\bf 94}, 053004 (2016).

\bibitem{mupb_terzi_2017}
M.~Terzi, S.~Tsysar, P.~Yuldashev, M.~Karzova, O.~Sapozhnikov, Generation of a
  vortex ultrasonic beam with a phase plate with an angular dependence of the
  thickness.
\newblock {\it Moscow Univ. Phys. Bull.\/} {\bf 72}, 61-67 (2017).

\bibitem{mssp_emanetoglu_1999}
N.~Emanetoglu, C.~Gorla, Y.~Liu, Y.~Lu, Epitaxial zno piezoelectric thin films
  for saw filters.
\newblock {\it Mat. Sci. Semiconductor Process.\/} {\bf 2}, 247-252 (1999).

\bibitem{piee_Booker_1950}
H.~Booker, P.~Clemmow, The concept of angular spectrum of plane waves, and its
  relation to that of polar diagram and aperture distribution.
\newblock {\it Proc. IEE - Part III: Radio \& Comm. Eng.\/} {\bf 97}, 11-17
  (1950).

\end{thebibliography}

\end{document}